\documentclass[a4paper,11pt]{article}
\usepackage{pos}
\usepackage{epstopdf}
\usepackage{here}
\usepackage{subcaption}
\usepackage{fancyhdr}
\usepackage{subcaption}
\usepackage{float}

\title{Performance studies for a next-generation optical sensor for IceCube-Gen2}
 \ShortTitle{Studies for an optical sensor for IceCube-Gen2}

\author{The IceCube-Gen2 Collaboration \vspace{2mm}}

\emailAdd{shimizu@hepburn.s.chiba-u.ac.jp}
\emailAdd{aya@hepburn.s.chiba-u.ac.jp}
\emailAdd{alexander.kappes@uni-muenster.de}

\abstract{We present performance studies of a segmented optical module for the IceCube-Gen2 detector.
 Based on the experience gained in sensor development for the IceCube Upgrade,
 the new sensor will consist of up to eighteen 4 inch PMTs housed in a transparent pressure vessel,
 providing homogeneous angular coverage. The use of custom molded
 optical `gel pads' around the PMTs enhances the photon capture
 rate via total internal reflection at the gel-air interface. This contribution
 presents simulation studies of various sensor, PMT, and gel pad geometries
  aimed at optimizing the sensitivity of the optical module in the face of confined space and harsh environmental conditions at the South Pole.

\vspace{4mm}
{\bfseries Corresponding authors:}
Nobuhiro Shimizu$^{1*}$, Aya Ishihara$^{1}$, Alexander Kappes$^{2}$\\
{$^{1}$ \itshape ICEHAP, Chiba University, Chiba, Japan}\\
{$^{2}$ \itshape Institut für Kernphysik, Westfälische Wilhelms-Universität Münster}\\[4mm]
$^*$ Presenter

\FullConference{37$^{\rm{th}}$ International Cosmic Ray Conference (ICRC 2021)\\
		July 12th -- 23rd, 2021\\
		Online -- Berlin, Germany}

}

\begin{document}
\maketitle

\section{IceCube, IceCube-Upgrade and IceCube-Gen2}

The IceCube Neutrino Observatory is a cubic kilometer neutrino telescope deployed in the glacial ice of the South Pole in Antarctica, and aims to observe high energy astrophysical neutrinos.
 Between depths of 1450~m to 2450~m, 5160 digital optical modules (DOMs) are
 deployed along 86 strings, and detect Cherenkov light emitted by secondary charged particles produced by the
 interactions of neutrinos. In the following text,
 this established project is referred to as {IceCube-Gen1}.
 
The IceCube Upgrade array, also known as IceCube-Gen2 Phase-1, consists of approximately 700 newly designed optical sensors and calibration modules, densely embedded near the bottom center of the existing IceCube Neutrino Observatory.  
The Upgrade array is currently planned to be deployed during 2022/2023 Austral Summer. 
 Two new optical module designs will be employed in the Upgrade: the 
 multi-PMT Digital Optical Module (mDOM)~\cite{citemDOM} and the Dual optical sensors in an Ellipsoid Glass for Gen2 (D-Egg)~\cite{citedEgg}, both designed to improve the photon detection efficiencies and the calibration capability of the detector. The IceCube Upgrade improves
 the reconstruction efficiency for few-GeV neutrinos, and will improve knowledge of the properties of scattering
 in ice, which yields better resolution on the neutrino events.
 
The IceCube upgrade is to be followed by the construction of the Gen2 array~\cite{IceCubeUpgrade}.
IceCube-Gen2 will consist of approximately 8~km$^3$ of
 instrumented ice and an array of approximately 10,000 optical sensors with a horizontal spacing of $\sim$240~m to achieve at least five times better sensitivity than that of IceCube for high energy neutrinos between 100~TeV to 1 PeV~\cite{gen2white}.

The different science targets goals of each project require different array configurations. Due to the horizontally sparse design of Gen2, maximising the photon collection efficiency of DOMs is particularly important. 

\section{Requirements and the baseline optical module designs for Gen2}
\begin{table}[b]
  \caption{Information on the various optical modules}
  \label{OM_table}
  \centering
  \begin{tabular}{cccccc}
	Name & PMT diameter & Number of& Glass diameter& Glass height \\ 
	 & [inch] & PMTs & [mm] & [mm] \\
	\hline \hline
  Gen1~DOM & 10 & 1 & 330 & 330 \\  
  mDOM & 3.15 & 24 & 356 & 411 \\
  D-Egg & 8 & 2 & 300 & 534 \\ 
  mEgg & 4 & 14 & 300 & 534 \\
  LOM$-{16}$ & 4 & 16 &  313 & 444\\ 
  LOM$-{18}$ & 4 & 18 &  305 & 540 \\ \hline \hline
  \end{tabular}\\
\end{table}
Gen2 will be a major construction effort. Optical sensors will be installed into water-filled holes in the South Pole ice made by a hot water drill, and  will be fixed in place when the hole ice re-freezes. For a significant cost saving as well as to complete
the construction within eight years, the drill hole size is being optimized.
This will limit the diameter of optical modules for Gen2 to $\lesssim12''$, smaller than that of Gen1 DOMs.
Regardless of their design, the modules will need to withstand the maximum pressure of 70~MPa
during the hole refreezing period, and 13-30~MPa will be the typical pressure after re-freezing.
Modules need to be sound against the thermal gradient of 20~$^\circ$C water during deployment
to -9~$^\circ$C to -40~$^\circ$C expected during operating conditions. 
Simulations indicate that optical modules with a three times higher effective area compoared to
Gen1 DOM and less than $\pm$~20\% variation in sensitivity across
the angle of photon incidence, will allow us to meet our science goals.

\begin{figure}[t]
 \begin{minipage}[b]{0.2\linewidth}
  \centering
  \includegraphics[keepaspectratio, width=3.5cm]  {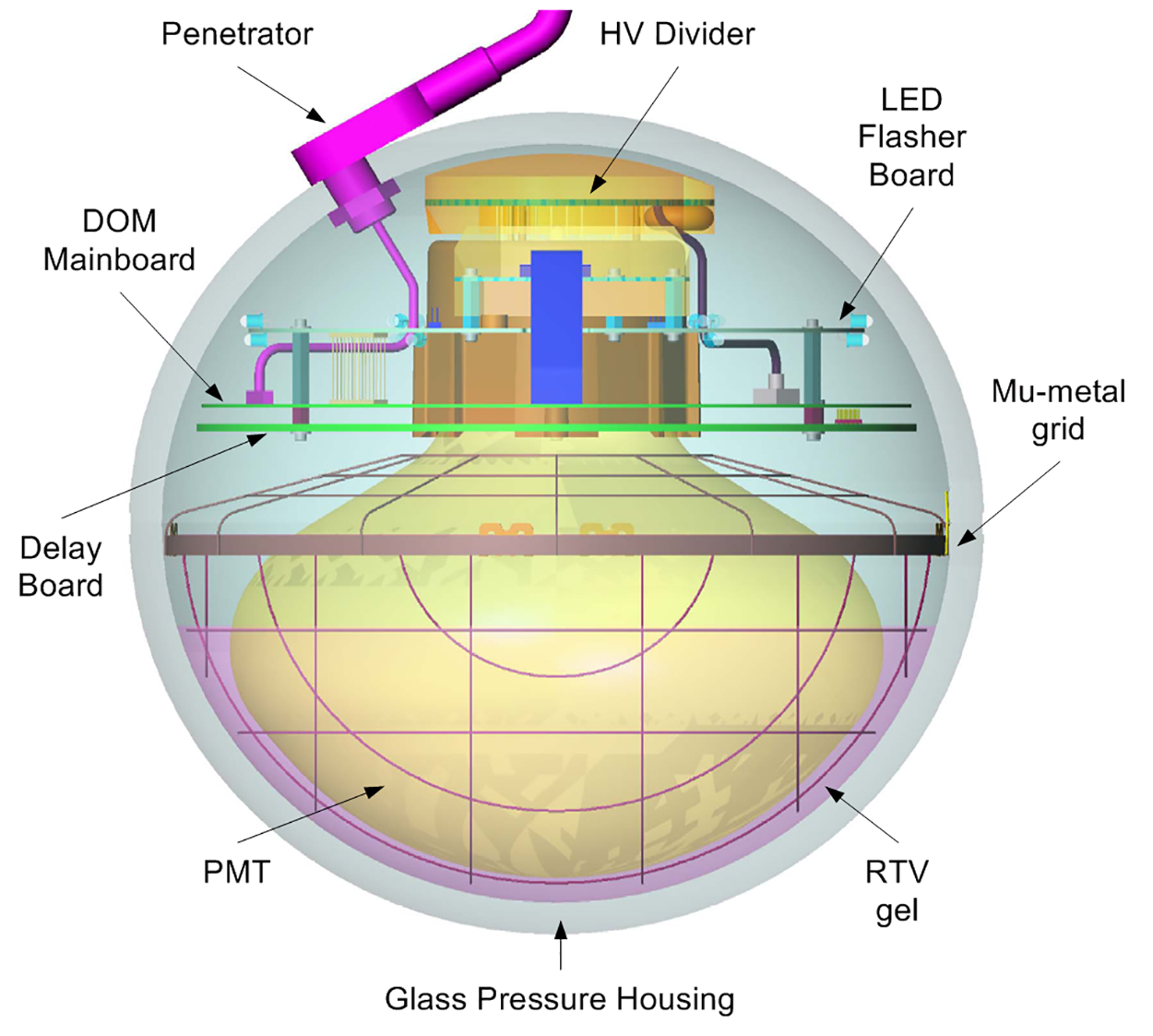}
  \subcaption{Gen1 DOM}\label{DOM}
  \end{minipage}
  \begin{minipage}[b]{0.2\linewidth}
  \centering
  \includegraphics[keepaspectratio, width=2.8cm]  {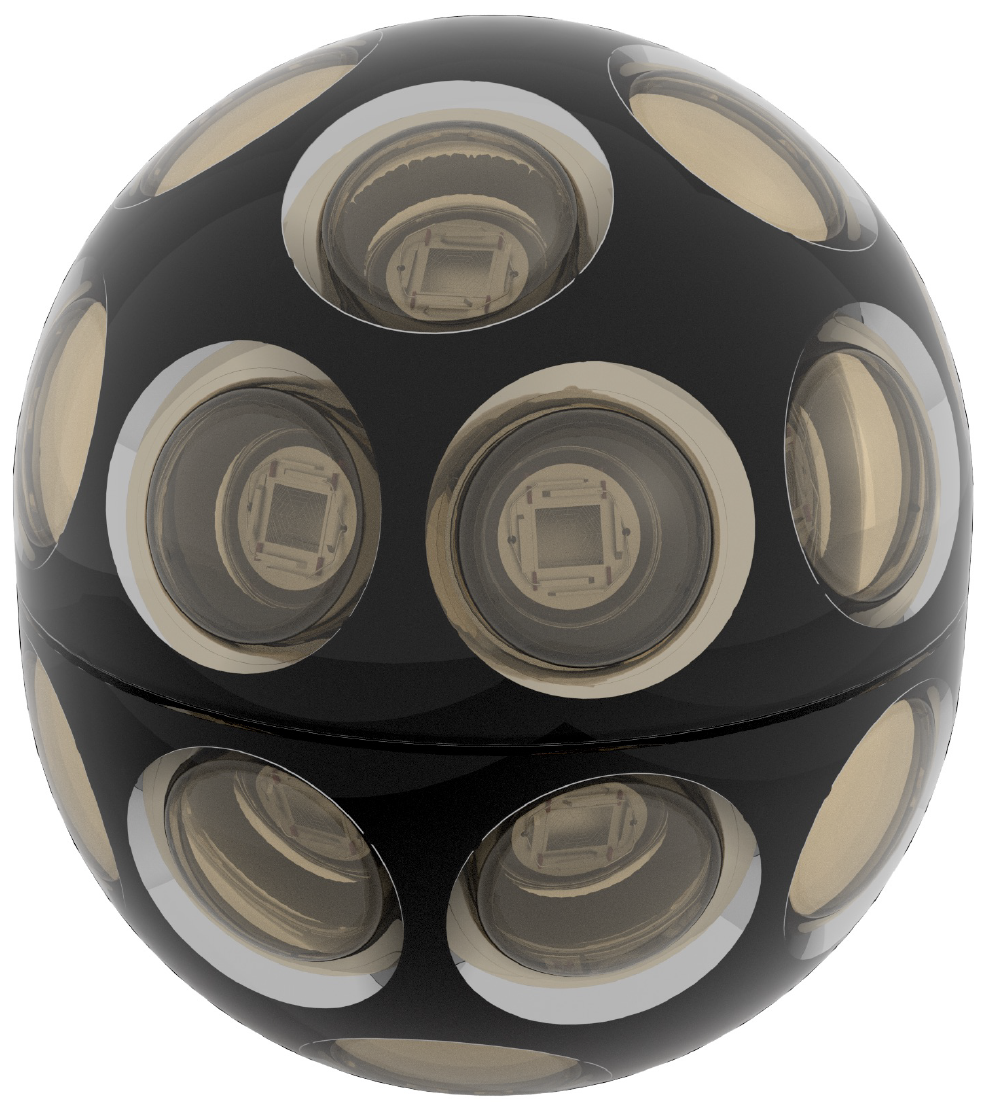}
  \subcaption{mDOM}\label{figmDOM}
 \end{minipage}
 \begin{minipage}[b]{0.2\linewidth}
  \centering
  \includegraphics[keepaspectratio, width=2.6cm]  {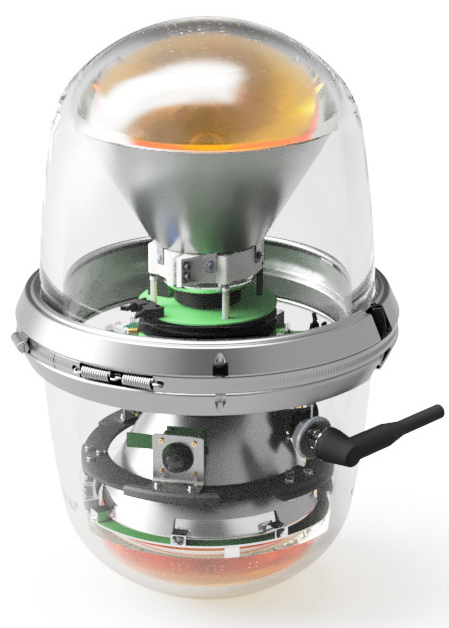}
  \subcaption{D-Egg}\label{figDEgg}
 \end{minipage}
 \begin{minipage}[b]{0.18\linewidth}
  \centering
  \vspace{-2mm}\hspace{-3mm} \includegraphics[keepaspectratio, width=2.9cm] {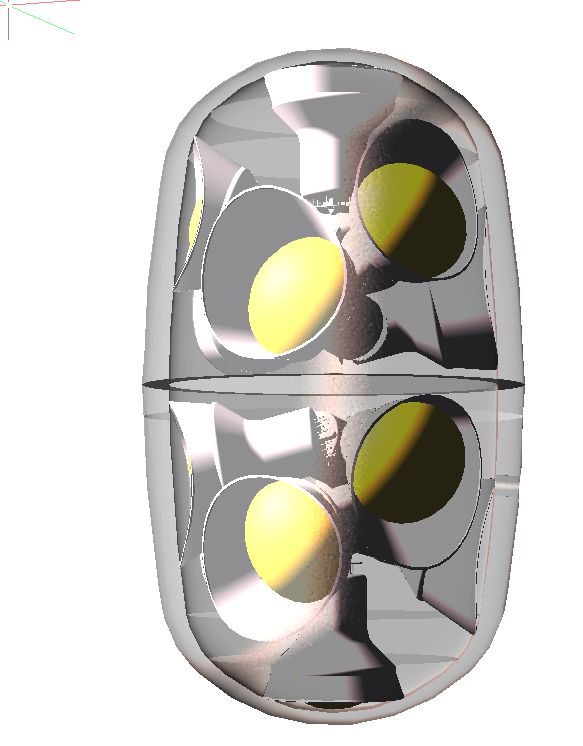}
  \subcaption{mEgg}\label{figLOM}
 \end{minipage}
 \begin{minipage}[b]{0.18\linewidth}
  \centering
  \includegraphics[keepaspectratio, width=2.4cm]  {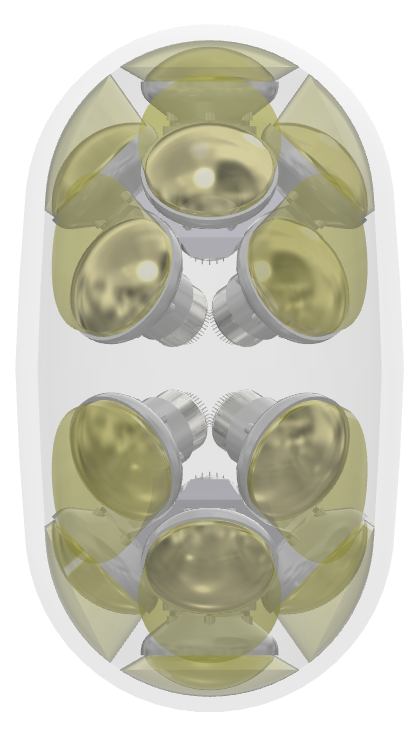}
  \subcaption{LOM$-{18}$}\label{poly20}
 \end{minipage}
 \caption{Schematic view of the optical modules. Gen-1 DOMs are used for the IceCube array, and mDOM and D-Eggs will be deployed in the IceCube-Upgrade holes. mEgg and LOM$-{18}$ are two designs for IceCube Gen2 studied here.}\label{om}
\end{figure}

Figure~\ref{om} shows the schematic view of the optical modules deployed as IceCube array (a), the modules to be installed in the Upgrade array (b,c) and the modules under investigation for the Gen2 array (d,e). The Gen1 DOM is a highly reliable optical sensor running for more than ten years with an extremely
small $(\leq0.5\%)$ post-deployment failure rate. A Gen1~DOM implements a single downward-facing 10~inch photomultiplier tube (PMT), the Hamamatsu R7081-2-MOD, and signals from the PMT are digitized in the readout board placed in the upper hemisphere of the module.
 
The mDOM for the IceCube Upgrade includes twenty-four 3.15 inch PMTs (Hamamatsu R15458-02) oriented in all directions, providing homogeneous sensitivity to photons. The side of the entrance window of the PMT is covered by an aluminum-coated reflector which improves the effective area by a factor of 20\%~\cite{citemDOM}.

The D-Egg, also being used in the IceCube Upgrade, implements two
 high-quantum efficiency (QE) 8 inch PMTs, the Hamamatsu R5912-100,
 in an elliptical glass vessel, facing both upward and downward. Because the top and bottom parts of the housing glass curvature are designed to be matched precisely with the R5912 PMT surface curvature of $\phi$131~mm, the actual photocathode area is 20\% more than when a flat disk of the same diameter is assumed. 
%

The Long Optical Module (LOM) for Gen2 \cite{Gen2LOM} is designed to fit into the smaller Gen2 holes while maximizing the effective photon sensitive area. The Gen2 hole diameter limits the diameter of the pressure vessel. The height of the housing is limited by the total weight of the module to be less than 26~kg. The optimized LOM modules include sixteen or eighteen 4~inch PMTs\footnote{Currently, both sixteen and eighteen-PMT designs (LOM$-{16}$ and LOM$-{18}$) are under investigation.}. The mEgg is another model of the optical module for Gen2 and uses fourteen 4~inch PMTs in the same glass housing as the D-Egg.

\section{Enlarging the effective photo-sensitive area with gel pad}


All existing modules (Gen1~DOM, mDOM, and D-Egg) utilize
ultraviolet (UV) transparent silicone elastomers (optical gel) for coupling between the glass vessel surface and PMT photocathodes. The transmittance of the glass and the optical gel is improved for the Upgrade modules compared to that of the Gen1~DOM~\cite{citedEgg, citedEgg2}.
For the Gen2 modules, because of the elongated vessel shape, the PMT surfaces are not vertically facing the glass surface. This complicates the module design. Either we have to use a large amount of optical gel fully filling the vessel or we need to come up with a technique to reduce the amount of optical gel, keeping good optical coupling properties.

\begin{figure}[h]
\begin{minipage}{0.5\hsize}
   \begin{center}
    \includegraphics[width=4.5cm]{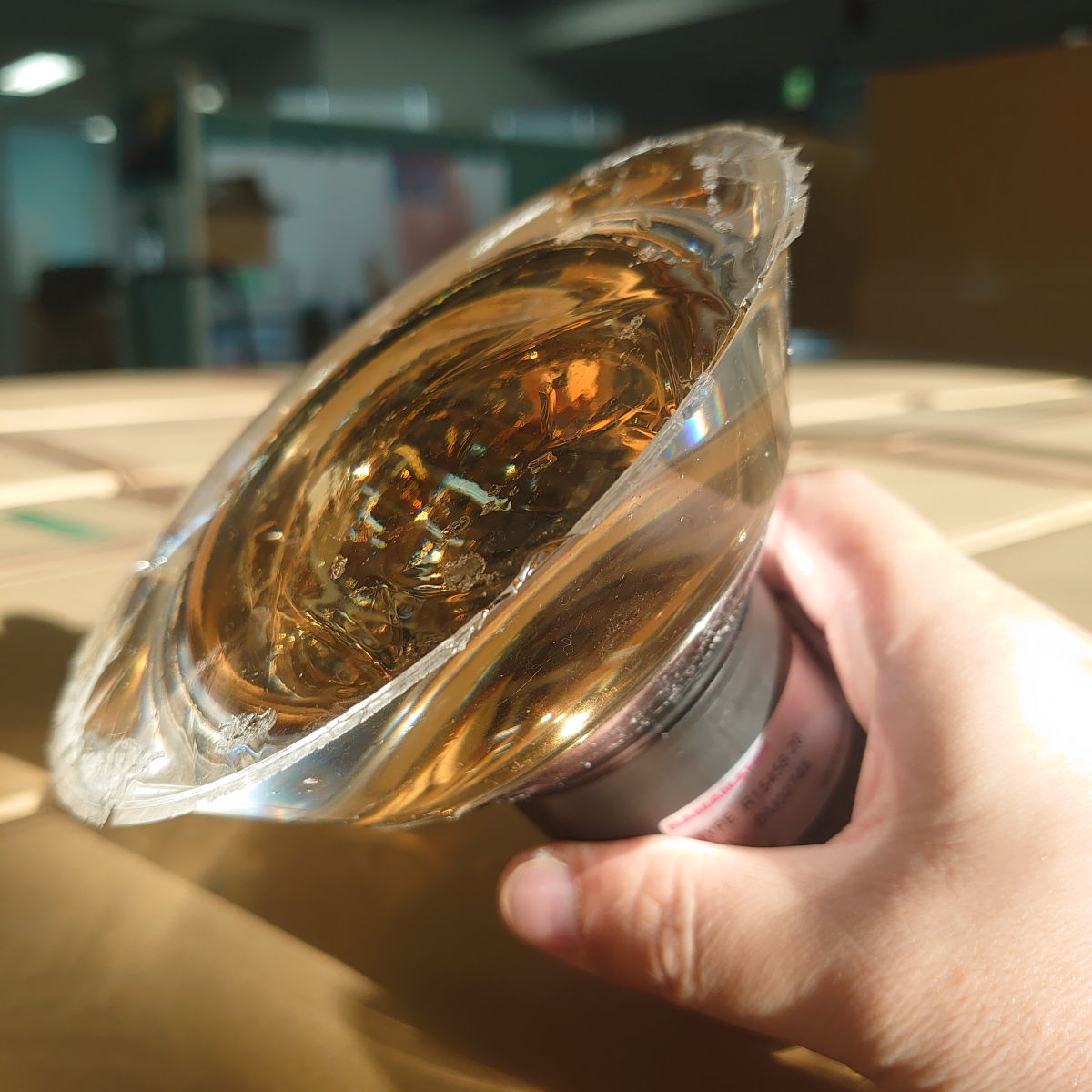} \label{gelpad_pic} \caption*{\hspace{0cm}(a)}
   \end{center}
 \end{minipage}
   \begin{minipage}{0.5\hsize}
     \begin{center}
     \includegraphics[width=7cm]{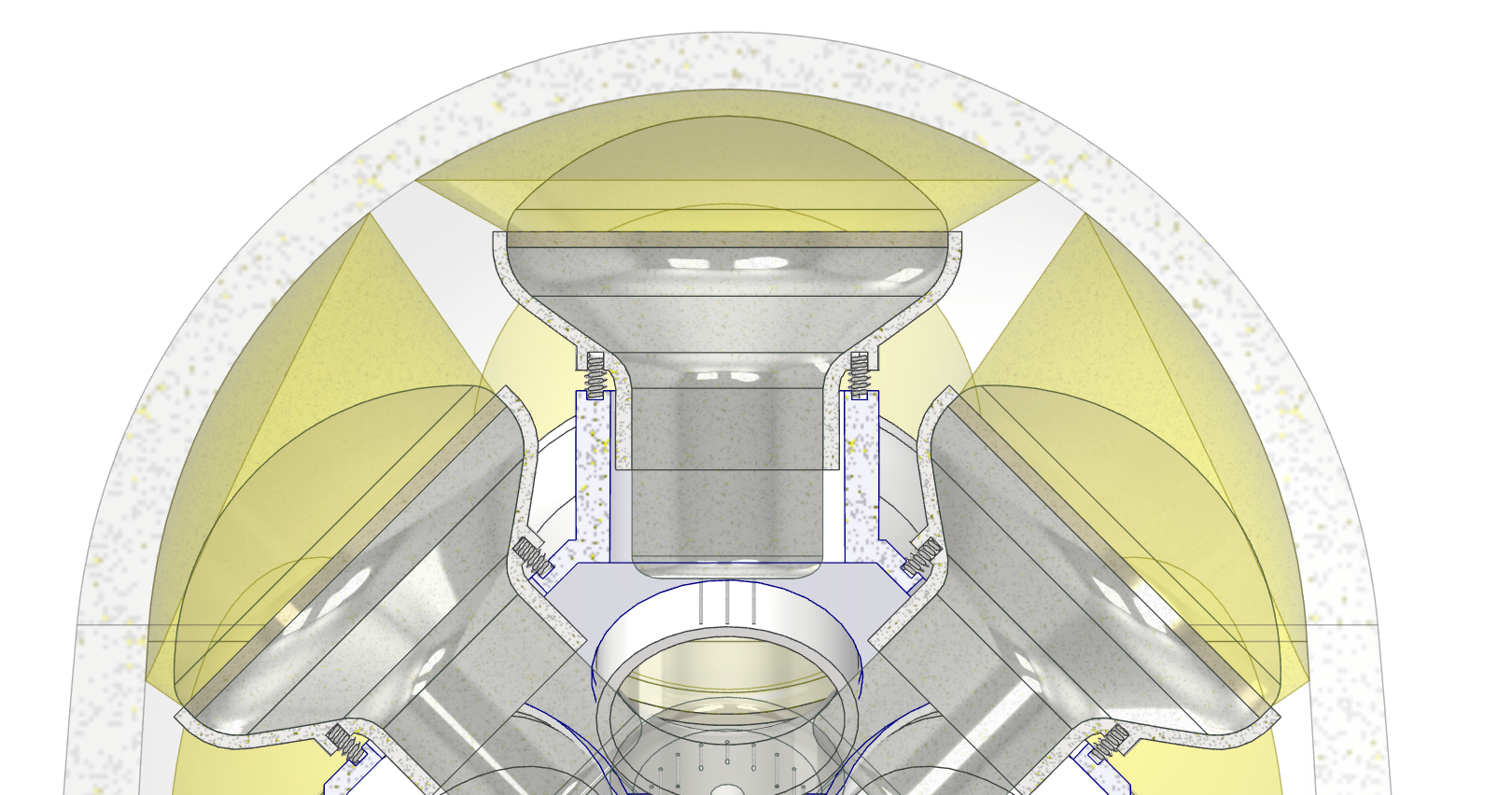} \label{gelpad_schematic} \caption*{\hspace{0cm}(b)}
     \end{center}
 \end{minipage}
 \caption{(a) Picture of a gel pad molded on a 3.15 inch PMT used for mDOM 
 (b) Schematic view of a tentative design of LOM$-{18}$.
 Yellow represents gel pads.} \label{gelpad} 
\end{figure}
In the mEgg and LOM, 
 the air gaps between PMTs and the pressure vessel are filled with
 an optical {\it gel pad}. 
Figures~\ref{gelpad}a and \ref{gelpad}b show a picture of the molded gel pad and
 its schematic view in the tentative mechanical design of the LOM.
 The side wall has a conical shape and works as a light collector thanks
 to the total internal reflection between the boundary of silicone and air.  

\begin{figure}[]
\begin{minipage}{0.8\hsize}
  \begin{center}
\includegraphics[width=9cm]{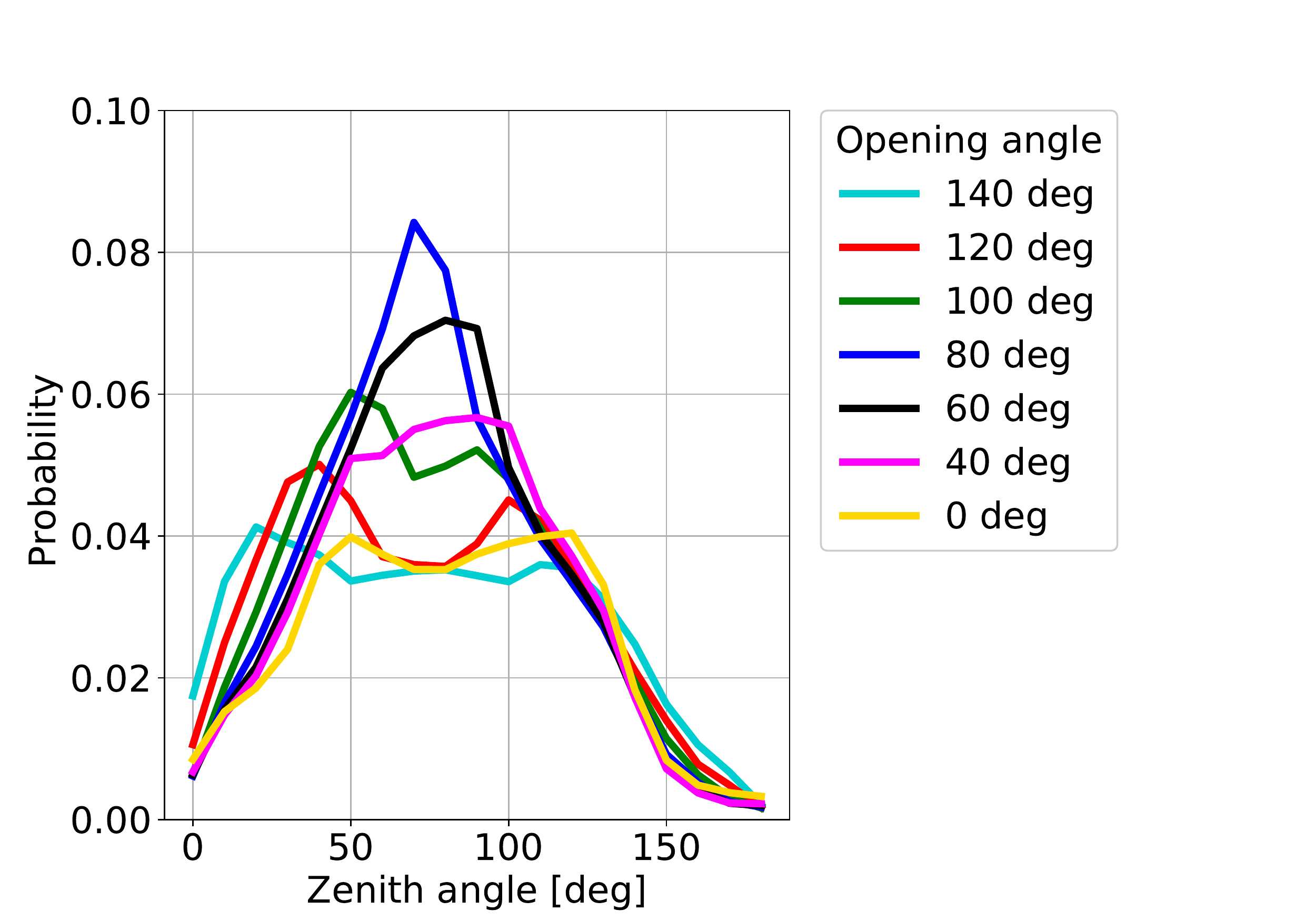}
\caption*{\hspace{-2.2cm}(a)}
\end{center}
 \end{minipage}
 \begin{minipage}{0.1\hsize}
 \begin{center}
  \vspace{0.2cm}\hspace{-1.5cm}\includegraphics[width=2.1cm]{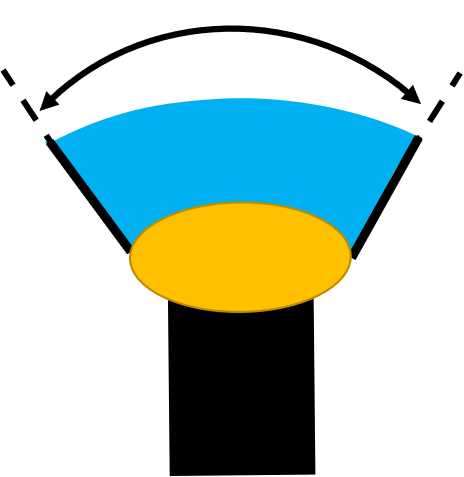}
  \caption*{\hspace{-13mm}(b)}
  \vspace{0.5cm} \hspace{-1.4cm} \includegraphics[width=2.8cm]{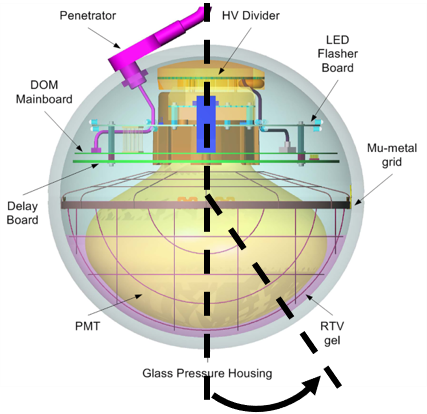}
  \caption*{\hspace{-13mm}(c)}
 \end{center}
 \end{minipage}
\caption{(a) Photon capture probability for various configurations of the opening angle as a function of the
 incident zenith angle of photons. The geometry of the mEgg was assumed.
  (b) Definition of the opening angle. 
  (c) Definition of the zenith angle.}
\label{gelpad_shape_optimization}
\end{figure}

To determine the optimal shape of the gel pad, we simulated the photon
 capture probability assuming the geometrical configuration of the mEgg.
 Using a GEANT4~\cite{citeGEANT4} based
 optical photon tracking simulator, 
 we injected a 380~nm parallel circular beam (the beam diameter was chosen to be
 large enough to fully cover the outer dimension of the mEgg)
 into the model of the mEgg, and calculated the probability that photons will arrive at any of the PMT photocathodes.
 Figure~\ref{gelpad_shape_optimization}a shows the
 evaluated photon capture probability for various values of the gel pad opening angle (its definition is shown in Fig.~\ref{gelpad_shape_optimization}b).
 After integration over the solid angle, we find that the 
 photon capture probability is maximum for
 opening angles from 60$^\circ$ to 80$^\circ$.

\begin{figure}[]
 \begin{minipage}{0.5\hsize}
  \begin{center}
   \includegraphics[width=.95\textwidth]{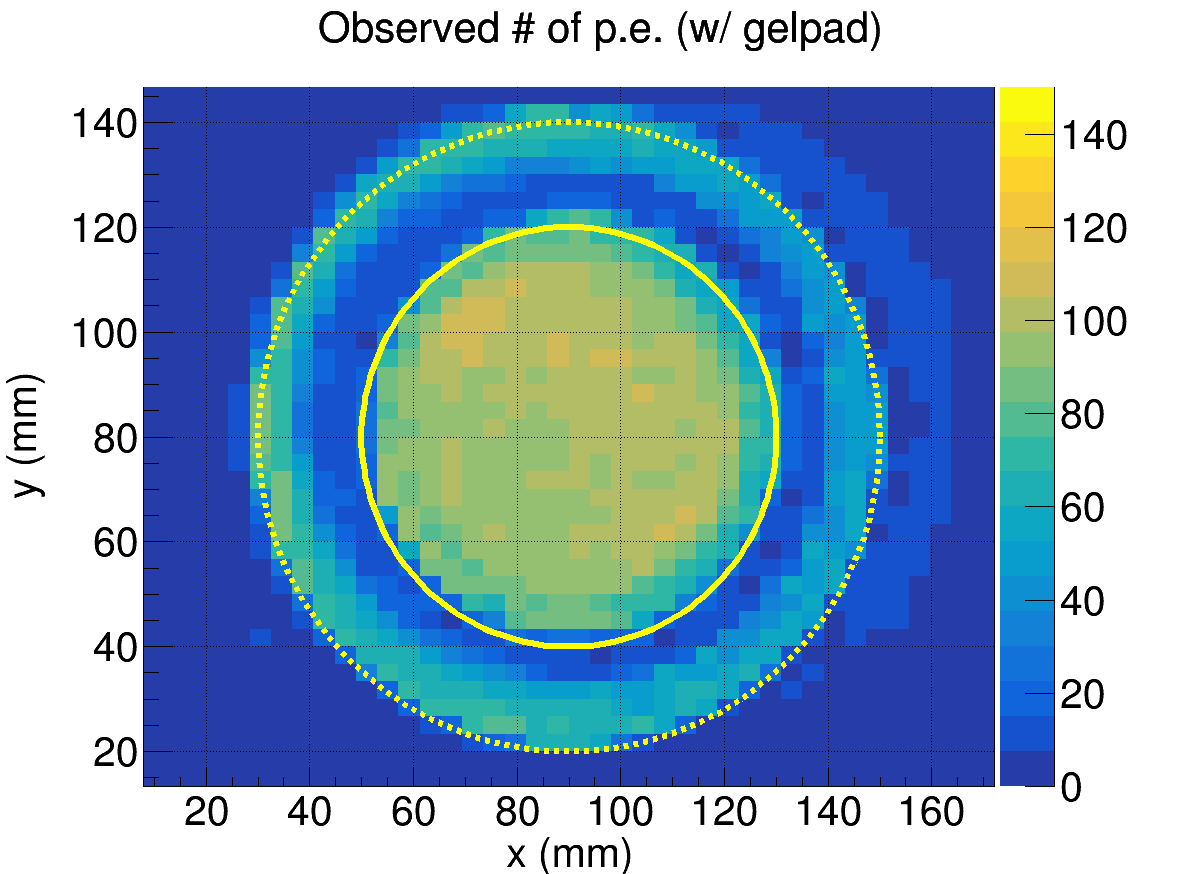}
  \end{center}
  \vspace{-0mm}\caption*{  (a)}
  \label{gelpad:one}
 \end{minipage}
 \begin{minipage}{0.5\hsize}
 \begin{center}
  \includegraphics[width=.95\textwidth]{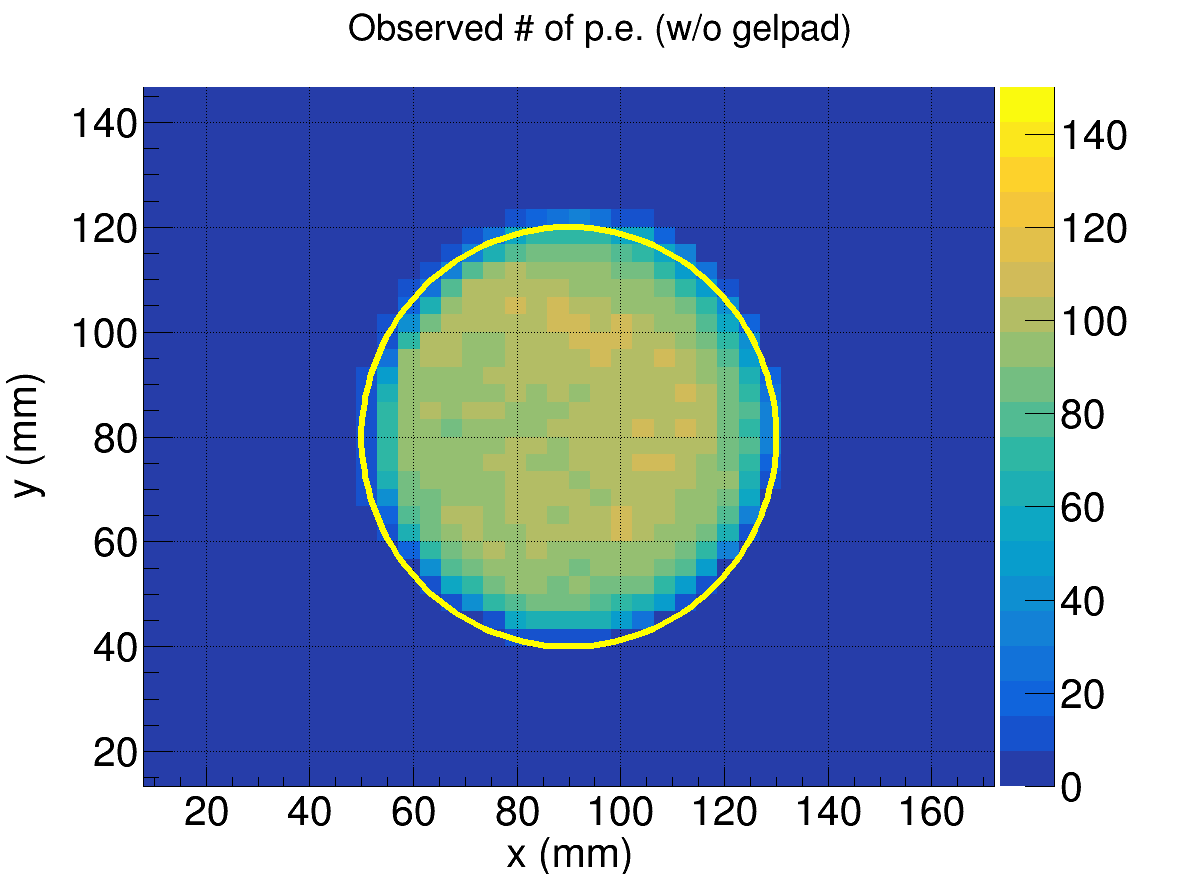}
 \end{center}
  \vspace{-0mm}\caption*{  (b)}
  \label{gelpad:two}
 \end{minipage}
 \caption{Distribution of the observed number of photoelectrons measured by a PMT (a) with and (b) without the gel pad. The solid and dashed lines
  represent the outer radii of the PMT and gel pad, respectively. }
 \label{eff_gelpad}
\end{figure}

The increase in photon capture probability with gel pads is due to
 the total reflection at their conical surface, and it is of importance to experimentally confirm the performance under realistic conditions in terms of control of surface conditions. We molded the gel pad onto a
 3.15 inch PMT (the one adopted for the mDOM) as shown in
 Fig.~\ref{gelpad}a)\footnote{At the time of writing, we did not have
 an electrically functional 4 inch PMT.} and investigated the feasibility of the gel pad scheme.
 We injected a parallel spot beam (wavelength 468~nm, spot diameter $\sim 2~\mathrm{mm}$) and counted the number of registered photons while scanning with the beam over the photocathode of the PMT.
 Figure~\ref{eff_gelpad} shows the distribution of the obtained number of photons
 with and without a gel pad. This confirmed that the gel pad indeed
 increases the sensitive
 region and causes no serious degradation of photon collection efficiency
 in the central region.
 For the parallel beam, an increase in effective area of 68\% was measured which is close to the 75\% estimated by the simulation.

\section{Simulation studies of Cherenkov photon sensitivities}

\subsection{Simulation setup}
\vspace{-5mm}
\begin{figure}[H]
\begin{center}
\begin{minipage}[b]{0.4\linewidth}
\hspace{8mm} \includegraphics[width=4cm]{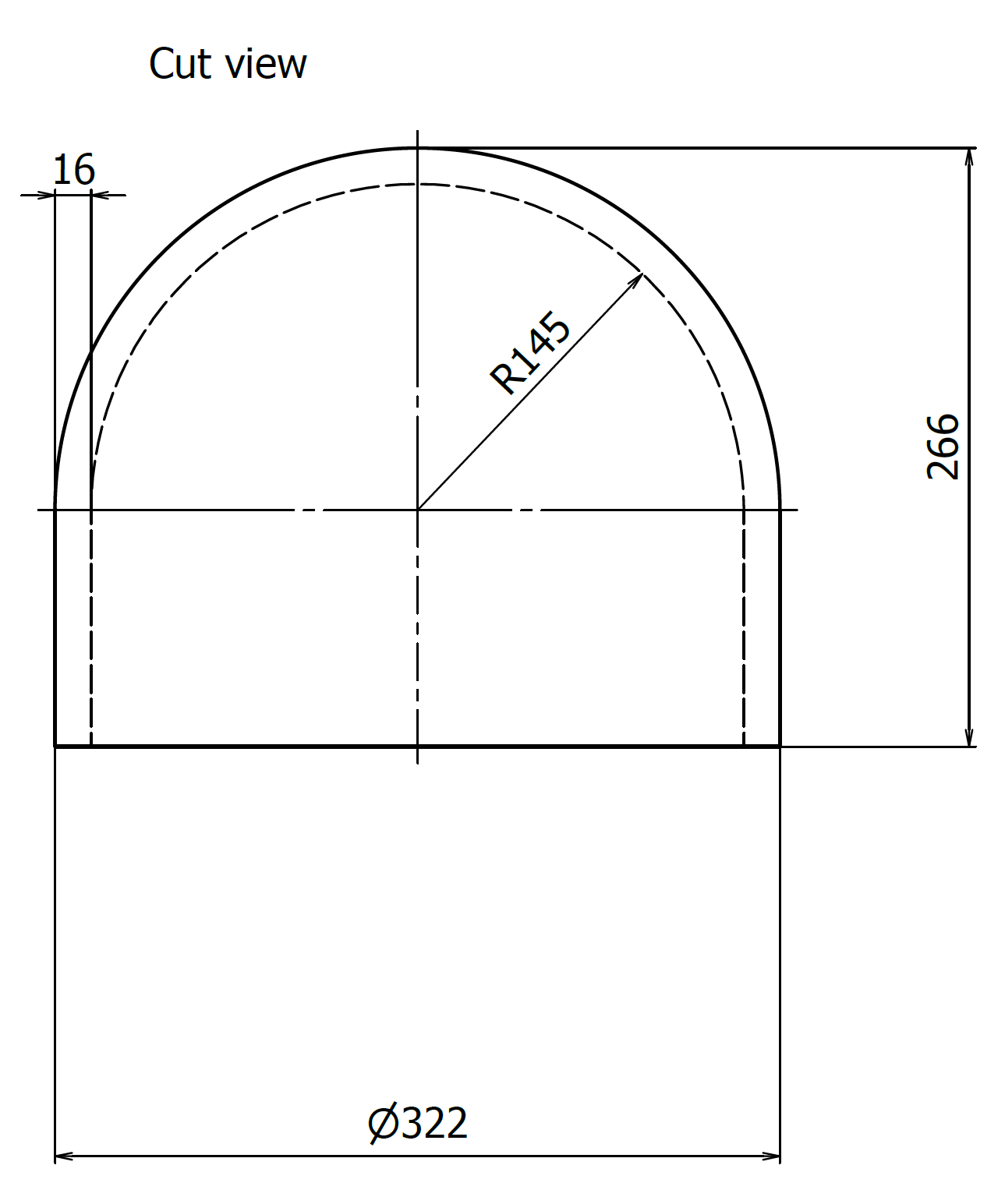} \caption*{(a)}
\end{minipage}
\begin{minipage}[b]{0.4\linewidth}
\hspace{1cm}\includegraphics[width=3.5cm]{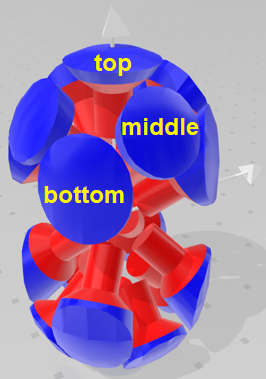}\caption*{(b)}
\end{minipage}
\caption{(a) Tentative design of the hemisphere of the LOM used in simulation.
(b) Implementation of the gel pads (blue) and PMTs (red)
in the glass. For a half hemisphere, there are three different levels and they are called bottom, middle, and top.  }\label{design_lom}
\end{center}
\end{figure}

Figure~\ref{design_lom} shows the preliminary design of the LOM$-{18}$ glass vessel, gel pads, and PMTs implemented in the simulation.
The transmittance of the glass and gel pad are assumed to be
the same as those of the D-Egg. The opening angles of the gel pads (see Fig.~\ref{gelpad_shape_optimization}b),
20$^\circ$, 10$^\circ$, and 70$^\circ$ (from the middle to top),
are chosen so as not to overlap with each other.
Eighteen 4~inch PMTs
are arranged and tilted in the glass as follows:
the top PMTs are vertically aligned,
the middle PMTs are tilted by 45$^\circ$, and
the bottom PMTs are tilted by 50$^\circ$.
In the current initial simulation studies, other components
such as internal structures and electrical boards are neglected,
though there is large enough space to fit them within the pressure vessel.

\subsection{Effective photo-sensitive area comparisons}

The performance of each optical module is evaluated using its effective area.
We simulated a parallel circular beam directed at the module, and
 calculated the effective area by 
\begin{align}
A(\theta, \phi, \lambda) = \frac{A_0}{N_\mathrm{gen}} \sum_{i: \mathrm{hit}} P(\lambda, \vec{r}_i),
\end{align}
where $\theta$ and $\phi$ are zenith and azimuth angles of the beam,
 $\lambda$ is the wavelength of the generated photons,
 $A_0$ is the area of the circular beam,
 $N_\mathrm{gen}$ is the number of generated photons,
 $P(\lambda, \vec{r_i})$ is the detection efficiency
 of the PMT for the wavelength of $\lambda$ 
 at the hit position of $i$-th trial,
 and the summation is over photons which hit on any photocathodes of PMTs. 

\begin{figure}[]
\begin{center}
\includegraphics[width=.6\textwidth]{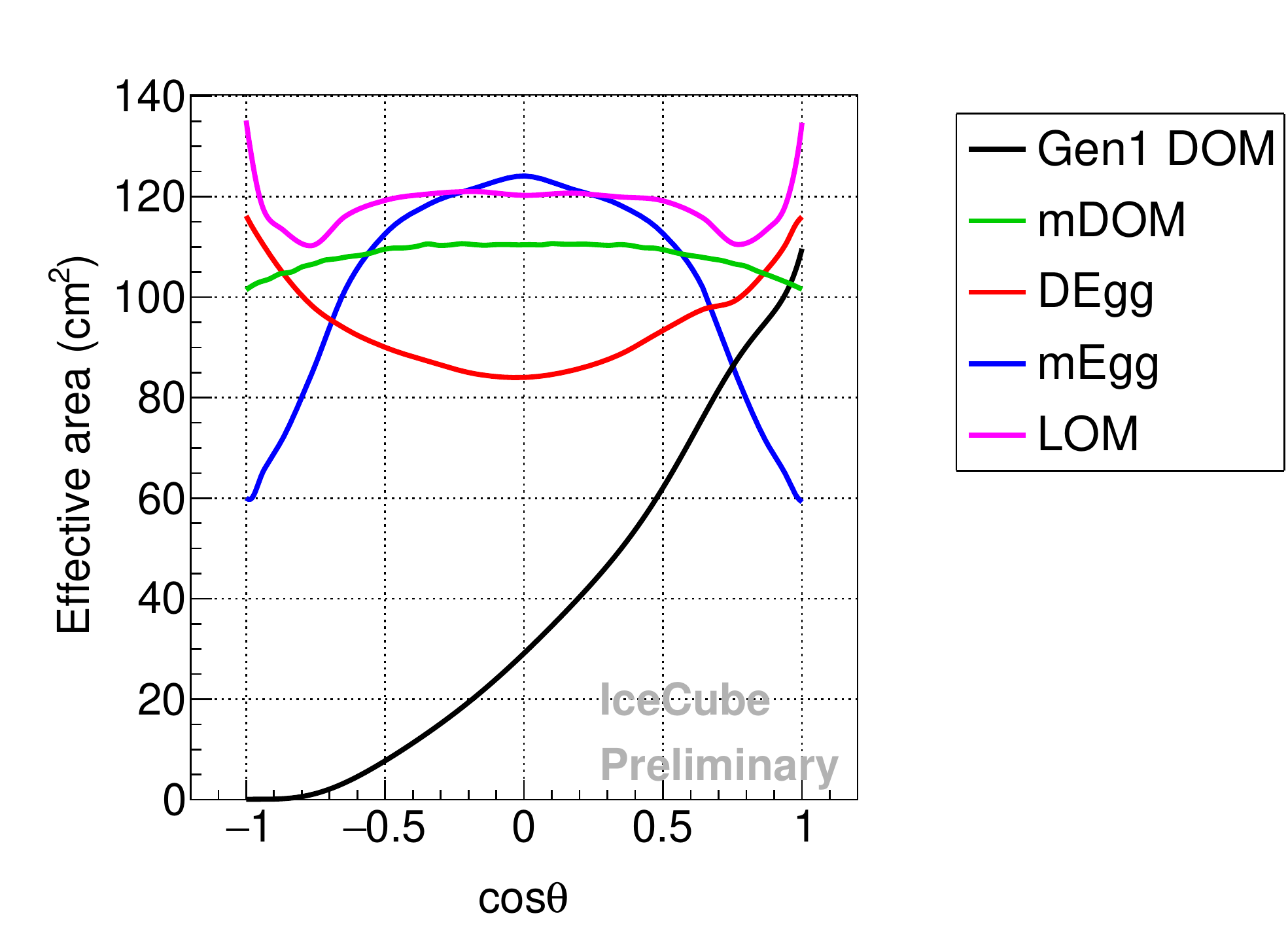}
\caption{Effective area at 400~nm as a function of cosine of zenith angle of incident photons. The downward-facing direction is defined as $\theta=0$ ($\cos{\theta}=1$).}
\label{eff}
\end{center}
\end{figure}

The comparison of the sensitivity of the individual modules is performed using the effective area at 400~nm. This is a typical wavelength of the incoming Cherenkov photons, since small wavelength photons are preferentially absorbed in the ice.
For small distances, the {\it Cherenkov-averaged} effective area can be another benchmark:
\begin{align}
\bar{A}(\theta, \phi) &= \frac{\displaystyle  \int_{270~\mathrm{nm}}^{700~\mathrm{nm}}  \mathrm{d} \lambda ~~A(\theta, \phi, \lambda)  P(\lambda) }{\displaystyle  \int_{270~\mathrm{nm}}^{700~\mathrm{nm}} \mathrm{d} \lambda ~~P(\lambda) } \\
P(\lambda) &= \frac{2\pi \alpha}{\lambda^2} \left(1-\frac{1}{\beta^2 n(\lambda)^2} \right),
\end{align}
where $P(\lambda)$ is the Cherenkov spectrum, $n$ is the refractive index of ice, and $\beta$ is the velocity
 of the traversing charged particle, assumed to be 1.
 Note that $P(\lambda)$ diverges for small wavelengths
 and $\bar{A}(\theta, \phi)$ depends on the lower cutoff of $\lambda $.

Figure~\ref{eff} shows the effective area of various optical modules as a function of cosine of the zenith angle, where $\theta=0$ is taken as the downward direction. In the calculations for the mEgg and LOM, we assumed that the 4~inch PMTs have the same quantum efficiency (QE) as the PMTs in the mDOM. 
The dependence of the detection efficiency on the hit position of the photons on the photocathode results from simulations performed by the manufacturer.
For the Gen1~DOM, D-Egg, and mDOM, we used experimentally measured values. By increasing the number of PMTs as well as the collection ability of photons through the use of gel pads, the mDOM and LOM show larger effective areas than the Gen1~DOM. The LOM design is characterized by a more homogeneous sensitivity to the direction of photon incidence than the mEgg.
 
Table~\ref{eff_area_comparison} summarizes the effective areas of the various optical modules
 after averaging over $\cos{\theta}$
 and $\phi$. The Cherenkov-averaged values are shown with respect to that of the Gen1~DOM.
 For both quantities, the eighteen-PMT model of the LOM will exceed
 the sensitivity of the Gen1~DOM by more than a factor three.
 Furthermore, if the same QE is available for the 4~inch PMTs as for the D-Egg PMTs, the sensitivity can be improved by another 30\%.

\begin{table}
\begin{center}

\caption{Effective area of optical modules (preliminary)}\label{eff_area_comparison}
\scalebox{0.9}{
\begin{tabular}{ccc}
	Name & Effective area (400~nm) & Cherenkov-averaged effective area ${}^*$ \\
	     & $[\mathrm{cm}^2]$ & [Ratio to Gen-1~DOM] \\
	\hline \hline
  Gen1 DOM  &  $34-37^{*}$  &  1  \\  
  mDOM &  108 & $3.5-4.0$ \\
  D-Egg & 94  & $2.8-3.2$ \\ 
  mEgg $ \dagger$&  103  & $3.2-3.6$  \\
  LOM$-{16}^\dagger$ & 105  & $3.2-3.7$ \\ 
  LOM$-{18}^\dagger$ & 118 & $3.6-4.2$  \\ \hline \hline
  \end{tabular}
  }\\
  \vspace{1mm} \hspace{-8mm} $*$~{\scriptsize Variation comes from a difference of the treatment of the detection efficiency of PMTs. }\\
  \vspace{-2mm}\hspace{-3cm} $ ~\dagger$~{\scriptsize \vspace{-3mm} If high quantum efficiency PMT is available, these improve by 30\%. 
  }

\end{center}
\end{table}

\section{Conclusion}

To improve the sensitivity to astrophysical neutrinos,
 IceCube-Gen2 will include approximately 10,000
 next-generation optical modules.
 To achieve the best performance at limited cost,
 the diameter of the optical modules is required to be $\sim12''$.
 The optimal design resulted in the concept of the Long Optical Module (LOM).
 The LOM will consist of sixteen or eighteen 4~inch PMTs,
 gel pads as optical coupler, and a transparent pressure vessel.
 A gel pad made of UV transparent silicone optical elastomer has a cone-shaped
 structure and increases
 the effective area via total internal reflection at its side surface.
 The concept of this scheme was experimentally confirmed using a parallel beam scan.
 To compare the effective area and its incident angular dependence
 for various optical modules,
 an optical photon simulation was performed.
 Compared to the Gen1 DOM, the LOM design will improve the effective
 area by more than a factor of three.
 The LOM shows high homogeneity with respect to the dependence of the sensitivity on the angle of photon incidence.
 Furthermore, if the same high-QE PMT as that of D-Egg is available in the future, the gain will further improve by an additional factor of 30\%.
 
\bibliographystyle{ICRC}
\bibliography{references}

%
%
%

\clearpage
\section*{Full Author List: IceCube-Gen2 Collaboration}

\scriptsize
\noindent
R. Abbasi$^{17}$,
M. Ackermann$^{71}$,
J. Adams$^{22}$,
J. A. Aguilar$^{12}$,
M. Ahlers$^{26}$,
M. Ahrens$^{60}$,
C. Alispach$^{32}$,
P. Allison$^{24,\: 25}$,
A. A. Alves Jr.$^{35}$,
N. M. Amin$^{50}$
% ...
\\
\\
% ...
$^{12}$ Universit{\'e} Libre de Bruxelles, Science Faculty CP230, B-1050 Brussels, Belgium \\
% ...
$^{17}$ Department of Physics, Loyola University Chicago, Chicago, IL 60660, USA \\
% ...
$^{22}$ Dept. of Physics and Astronomy, University of Canterbury, Private Bag 4800, Christchurch, New Zealand \\
% ...
$^{24}$ Dept. of Astronomy, Ohio State University, Columbus, OH 43210, USA \\
$^{25}$ Dept. of Physics and Center for Cosmology and Astro-Particle Physics, Ohio State University, Columbus, OH 43210, USA \\
$^{26}$ Niels Bohr Institute, University of Copenhagen, DK-2100 Copenhagen, Denmark \\
% ...
$^{32}$ D{\'e}partement de physique nucl{\'e}aire et corpusculaire, Universit{\'e} de Gen{\`e}ve, CH-1211 Gen{\`e}ve, Switzerland \\
% ...
$^{35}$ Karlsruhe Institute of Technology, Institute for Astroparticle Physics, D-76021 Karlsruhe, Germany  \\
% ...
$^{50}$ Bartol Research Institute and Dept. of Physics and Astronomy, University of Delaware, Newark, DE 19716, USA \\
% ...
$^{60}$ Oskar Klein Centre and Dept. of Physics, Stockholm University, SE-10691 Stockholm, Sweden \\
% ...
$^{71}$ DESY, D-15738 Zeuthen, Germany \\
% ...

\subsection*{Acknowledgements}

\noindent
USA {\textendash} U.S. National Science Foundation-Office of Polar Programs,
U.S. National Science Foundation-Physics Division,
U.S. National Science Foundation-EPSCoR,
Wisconsin Alumni Research Foundation,
Center for High Throughput Computing (CHTC) at the University of Wisconsin{\textendash}Madison,
Open Science Grid (OSG),
Extreme Science and Engineering Discovery Environment (XSEDE),
Frontera computing project at the Texas Advanced Computing Center,
U.S. Department of Energy-National Energy Research Scientific Computing Center,
Particle astrophysics research computing center at the University of Maryland,
Institute for Cyber-Enabled Research at Michigan State University,
and Astroparticle physics computational facility at Marquette University;
Belgium {\textendash} Funds for Scientific Research (FRS-FNRS and FWO),
FWO Odysseus and Big Science programmes,
and Belgian Federal Science Policy Office (Belspo);
Germany {\textendash} Bundesministerium f{\"u}r Bildung und Forschung (BMBF),
Deutsche Forschungsgemeinschaft (DFG),
Helmholtz Alliance for Astroparticle Physics (HAP),
Initiative and Networking Fund of the Helmholtz Association,
Deutsches Elektronen Synchrotron (DESY),
and High Performance Computing cluster of the RWTH Aachen;
Sweden {\textendash} Swedish Research Council,
Swedish Polar Research Secretariat,
Swedish National Infrastructure for Computing (SNIC),
and Knut and Alice Wallenberg Foundation;
Australia {\textendash} Australian Research Council;
Canada {\textendash} Natural Sciences and Engineering Research Council of Canada,
Calcul Qu{\'e}bec, Compute Ontario, Canada Foundation for Innovation, WestGrid, and Compute Canada;
Denmark {\textendash} Villum Fonden and Carlsberg Foundation;
New Zealand {\textendash} Marsden Fund;
Japan {\textendash} Japan Society for Promotion of Science (JSPS)
and Institute for Global Prominent Research (IGPR) of Chiba University;
Korea {\textendash} National Research Foundation of Korea (NRF);
Switzerland {\textendash} Swiss National Science Foundation (SNSF);
United Kingdom {\textendash} Department of Physics, University of Oxford.

\end{document}